\documentclass[aps,prl,reprint,superscriptaddress]{revtex4-1}

\usepackage[dvips]{graphicx}

\begin{document}


\title{Critical Magnetic Field Strength for Suppression of the
  Richtmyer-Meshkov Instability in Plasmas} 


\author{Takayoshi Sano}
\email{sano@ile.osaka-u.ac.jp}
\affiliation{Institute of Laser Engineering, Osaka University, Suita,
  Osaka 565-0871, Japan} 

\author{Tsuyoshi Inoue}
\affiliation{Department of Physics and Mathematics, Aoyama Gakuin
  University, Sagamihara, Kanagawa 252-5258, Japan}

\author{Katsunobu Nishihara}
\affiliation{Institute of Laser Engineering, Osaka University, Suita,
  Osaka 565-0871, Japan} 


\date{\today}

\begin{abstract}
The critical strength of a magnetic field required for the suppression
of the Richtmyer-Meshkov instability (RMI) is investigated numerically
by using a two-dimensional single-mode analysis.  
For the cases of MHD parallel shocks, the RMI can be stabilized as a
result of the extraction of vorticity from the interface.  
A useful formula describing a critical condition for MHD RMI has been
introduced, and which is successfully confirmed by the direct
numerical simulations. 
The critical field strength is found to be largely depending on the
Mach number of the incident shock.
If the shock is strong enough, even low-$\beta$ plasmas can be subject
to the growth of the RMI.  
%
\end{abstract}


\maketitle



The Richtmyer-Meshkov instability (RMI) in magnetohydrodynamics is of
great interest in many fields such as astrophysical phenomena,
laboratory experiments, and inertial confinement fusion
\cite{brouillette02,nishihara10}.   
The RMI occurs when an incident shock strikes a corrugated contact
discontinuity \citep{richtmyer60,meshkov69}. 
A strong shock wave traveling through the density inhomogeneity of
magnetized interstellar medium is a promising site of the RMI.
This astrophysically common event plays a key role to determine the
dynamics of supernova remnants \cite{inoue12a} and gamma ray bursts
\cite{inoue11}. 
Recent laboratory experiments are designed to test the magnetic field
amplification due to the RMI by the use of laser-induced
shock waves \cite{kuramitsu11}.
In inertial confinement fusion, the RMI excited at several capsule
interfaces amplifies the perturbations that seed the Rayleigh-Taylor
instability.
For the fast ignition approach, the utilization of an external
magnetic field to guide the fast electrons is discussed proactively,
and which shed the light on the impact of MHD instabilities during the
implosion \cite{chang11,strozzi12}. 

Inclusion of a magnetic field brings two important consequences into
the RMI, which are the amplification of an ambient field and the
suppression of the unstable motions.  
The magnetic field can be amplified by the stretching motions at the
interface associated with the RMI \cite{sano12}.
\citet{samtaney03} have shown that a strong magnetic field inhibits
the nonlinear turbulent motions of the RMI. 
The vorticity generated by the interaction between a shock front and a
corrugated contact discontinuity is the driving mechanism for the RMI. 
For the cases of MHD parallel shocks, the role of the magnetic field
is to prevent the deposition of the vorticity on the interface, and
stabilize the RMI \citep{wheatley05,wheatley09}.  

In the weakest field limit, the RMI should happen just like the
hydrodynamical cases, so that there must exist the critical field
strength for the suppression. 
However, how large field is necessary to kill the RMI is still an open
question.
The previous works are mostly focusing on the weak shock cases.
The Mach number of astrophysical and laboratory shocks takes various
values including extremely large ones.
Thus, the goal of this letter is to evaluate the critical field
strength by studying the evolutions of the RMI systematically in a quite
wide range of parameters. 

\begin{figure}
\includegraphics[scale=0.63,clip]{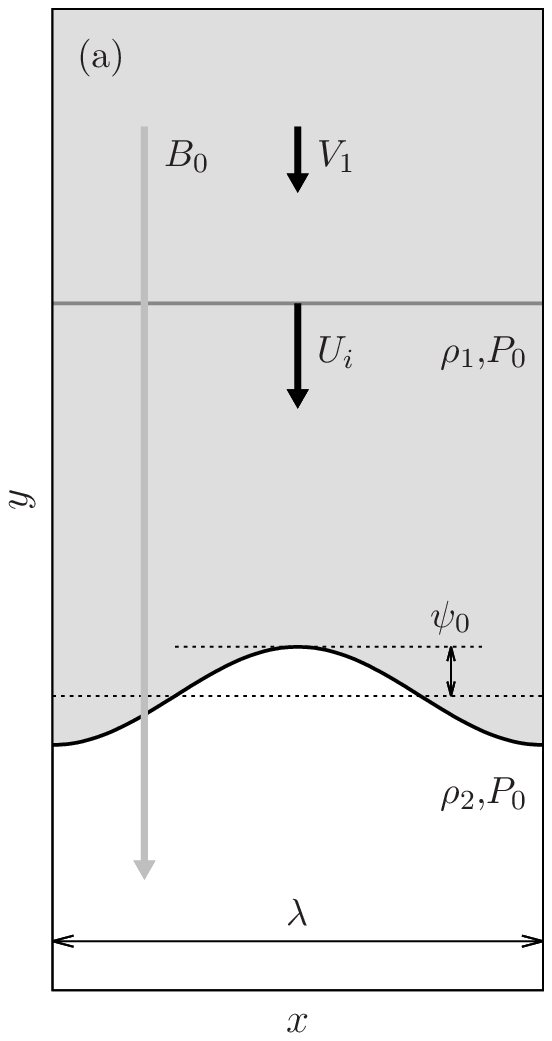}%
\hspace{0.2cm}{\includegraphics[scale=0.45,clip]{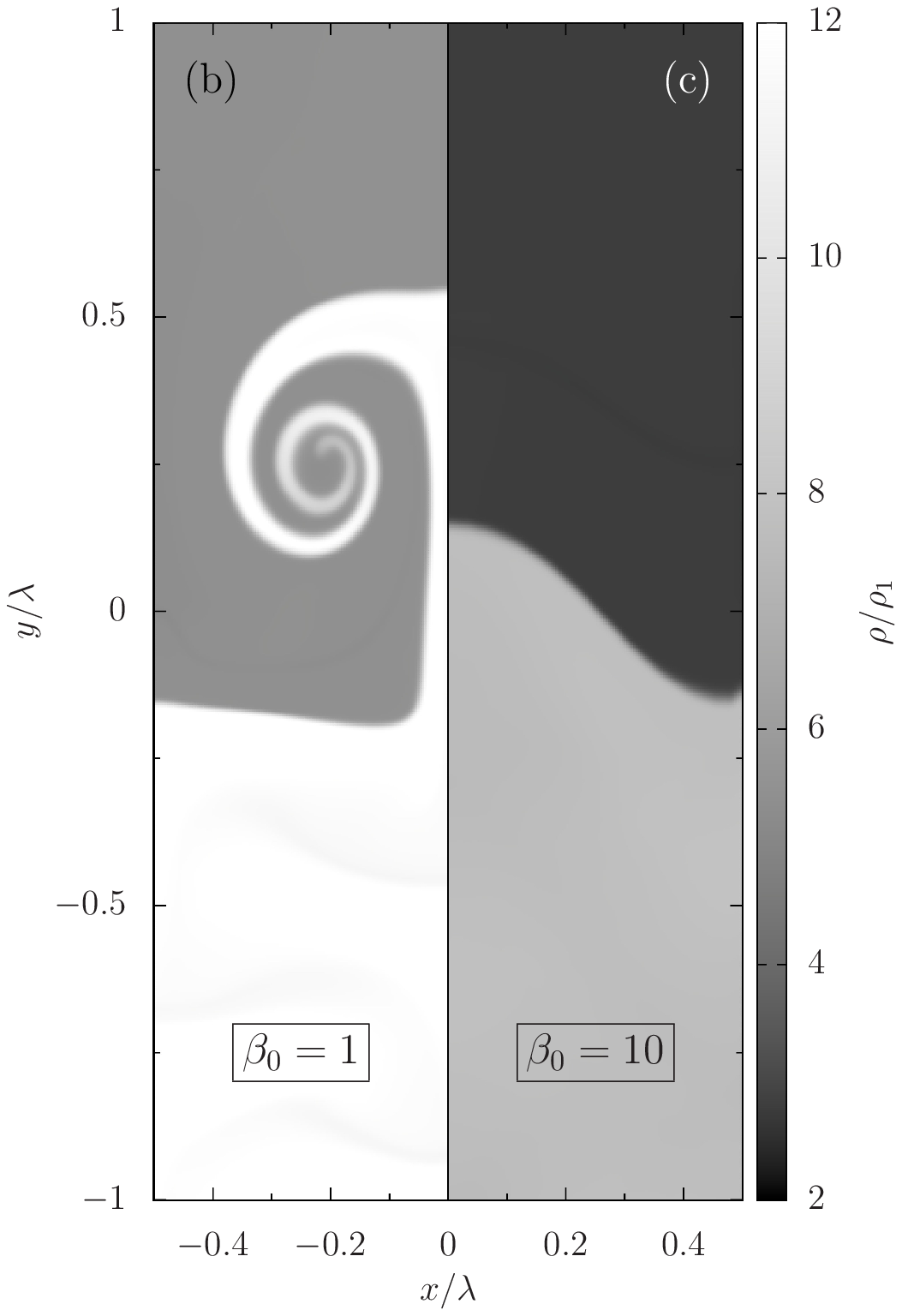}}
\caption{
(a) 
Sketch of the initial configuration for the single-mode RMI simulations. 
The sinusoidal corrugation of a contact discontinuity is given by $y =
\psi_0 \cos (k x)$ and an incident shock moves in the $y$ direction
with the velocity $U_i (<0)$. 
(b,c)
Numerical results of the RMI for the models with (b) a stronger field
$\beta_0 = 1$ and (c) a weaker field $\beta_0 = 10$. 
The gray color denotes the density profiles at the later evolutionary
stage of the RMI taken at $k v_{\rm lin} t = 10$.
For the both models, the initial corrugation amplitude is
$\psi_0/\lambda = 0.1$ and the initial density jump is $\rho_2 / \rho_1 = 3$.
The Mach number of the incident shock is (b) $M = 200$ and (c) $M =
2$. 
\label{fig1}}
\end{figure}

We adopted a single-mode analysis for the MHD RMI in two-dimensions, 
the same as \citet{sano12} in which the detailed settings and 
numerical method are described.   
The initial configuration of the system is illustrated in
Fig.~\ref{fig1}(a). 
The contact discontinuity separates two fluids with the densities
$\rho_1$ and $\rho_2 (> \rho_1)$.
The corrugation of the contact surface is a key ingredient for
the RMI.
The interface is assumed to be sinusoidal with a wavelength $\lambda =
2\pi/k$ where $k$ is the wavenumber.  
Then, the discontinuity is characterized by two parameters; the ratio
of the corrugation amplitude to the wavelength $\psi_0/\lambda$ and
the density jump $\rho_2/\rho_1$.

The incident shock propagates through the light fluid 1 with a shock
velocity $U_i$ and hits the interface at $t = 0$, where $V_1$ is the
fluid velocity behind the shock [see Fig.~\ref{fig1}(a)]. 
The shock strength is indicated by the sonic Mach number $M =
|U_i|/c_{s1}$, where $c_{s1} = (\gamma P_0/\rho_1)^{1/2}$ is the sound
speed in the fluid 1, $\gamma$ is the ratio of the specific heats, and
$P_0$ is the initial pressure. 
When the incident shock hits the corrugated contact discontinuity, the
reflected and transmitted shocks start to travel from the interface in
the opposite directions with the velocities $U_r$ and $U_t$,
respectively. 
The RMI can take place in such the situation, and a spike grows
linearly with time from the heavy fluid toward the light one. 

For the hydrodynamical cases, the asymptotic growth velocity of the
RMI can be derived from the linear analysis
\citep{wouchuk96,wouchuk97}, which is written as 
\begin{equation}
v_{\rm lin} = 
\frac{\rho_1^{\ast} \delta v_1^{\ast} - \rho_2^{\ast} \delta v_2^{\ast}}
{\rho_1^{\ast} + \rho_2^{\ast}} \;,
\label{eq:vlin}
\end{equation}
where $\rho_1^{\ast}$ and $\rho_2^{\ast}$ are the postshock densities
of each fluid, and $\delta v_1^{\ast} = k \psi_r (v^{\ast} - V_1)$ and
$\delta v_2^{\ast} = k \psi_t v^{\ast}$ are the tangential velocities
generated by the refraction of fluid motions at the reflected and
transmitted shocks. 
Here, $v^{\ast}$ is the postshock velocity of the interface, and
$\psi_r = \psi_0 (1 - U_r/U_i)$ and $\psi_t = \psi_0 (1 - U_t/U_i)$
are the initial ripple amplitudes of each shock.  
The Atwood number of the postshocked interface can be defined by
$A^{\ast} = (\rho_2^{\ast} - \rho_1^{\ast})/(\rho_2^{\ast} +
\rho_1^{\ast})$.  
The growth velocity $v_{\rm lin}$ given by Eq.~(\ref{eq:vlin}) is
an exact solution in the weak shock limit.
When the Mach number is large, the bulk vorticity left behind the
rippled transmitted shock reduces $v_{\rm lin}$ by a factor of a few
\cite{wouchuk97}.  
The MHD effects could also modify the growth velocity
\cite{wheatley05,wheatley09}. 
However, as with our previous analysis \cite{sano12}, we adopt
Equation~(\ref{eq:vlin}) for the typical growth velocity.   

In this letter, only the parallel shock cases are considered, so that
a uniform magnetic field perpendicular to the shock surface, $(B_x,
B_y) = (0, B_0)$, is assumed. 
The initial field strength is given by the plasma beta $\beta_0 = 8
\pi P_0/B_0^2$ in the preshocked regions. 
We solve the ideal MHD equations, and most of the calculations use a
grid resolution of $\Delta_x$ = $\Delta_y$ = $\lambda/256$ unless
otherwise stated.  

Figures~\ref{fig1}(b) and \ref{fig1}(c) show the simulation results of
the density distribution at the later stage of the RMI ($k v_{\rm lin}
t = 10$) for two different models.  
The mushroom-shaped spike and roll-up due to the growth of the RMI can
be seen in Fig.~\ref{fig1}(b), while the corrugation amplitude does
not change by much for the model in Fig.~\ref{fig1}(c). 
The plasma beta is initially $\beta_0 = 1$ and 10 for the models in
Figs.~\ref{fig1}(b) and \ref{fig1}(c), respectively, and then 
the field strength is stronger in the unstable model and weaker in
the stabilized model. 

Interestingly, even when the preshocked plasma is strongly magnetized
as $\beta_0 = 1$, the RMI is not necessarily suppressed. 
Reversely, there exists a case that the RMI is quenched completely by
a weaker field with $\beta_0 = 10$.
These results clearly demonstrate that the critical field strength
cannot be simply described as $\beta \sim 1$.  
Then, what determines the conditions for suppression of the RMI?

It is found that the Mach number of the incident shock has a huge impact
on the critical field strength.
For the both models in Fig.~\ref{fig1}, the parameters related to the
contact discontinuity are identical, that is, $\psi_0 / \lambda = 0.1$
and $\rho_2 / \rho_1 = 3$. 
But, the Mach number is $M = 200$ and 2 in Figs.~\ref{fig1}(b) and
\ref{fig1}(c), respectively, which indicates that the weaker shock
case could be stabilized by a much weaker field.

\citet{wheatley09} have found that the presence of a magnetic field
affects the refraction of fluid motions at the reflected and transmitted
shocks.
For the cases of MHD, each shock could split into a combination of the
waves and/or discontinuities associated with the fast, Alfv{\'e}n, and
slow modes. 
In contrast to hydrodynamical shocks, the jump in the
tangential velocity can exist only at the MHD waves, and is not
allowed to be at the contact discontinuity. 

Let us consider the conservation laws related to the momentum and the
tangential electric field component in the discontinuity frame 
\cite{landau60}; 
\begin{eqnarray}
\left[ \rho v_n v_t - \frac{B_n B_t}{4 \pi} \right] &=& 0 \;,\\
\left[ v_n B_t - v_t B_n \right] &=& 0 \;,
\end{eqnarray}
where the square brackets mean the difference between the values on
the two sides of the discontinuity.  
The subscript $n$ and $t$ denote the normal and transverse components,
respectively.  
For the contact discontinuity, the normal velocity is $v_n = 0$, so
that it should be satisfied that 
$[v_t] = [B_t] = 0$  when $B_n \neq 0$. 
Therefore, the vortex sheet and the current sheet cannot be located
at the contact discontinuity.
Although the vorticity is generated instantaneously at the interface,
it must move away from the interface with the MHD waves. 
This feature could affect seriously on the nonlinear evolutions of the
MHD RMI.  

\begin{figure}
\includegraphics[scale=0.54,clip]{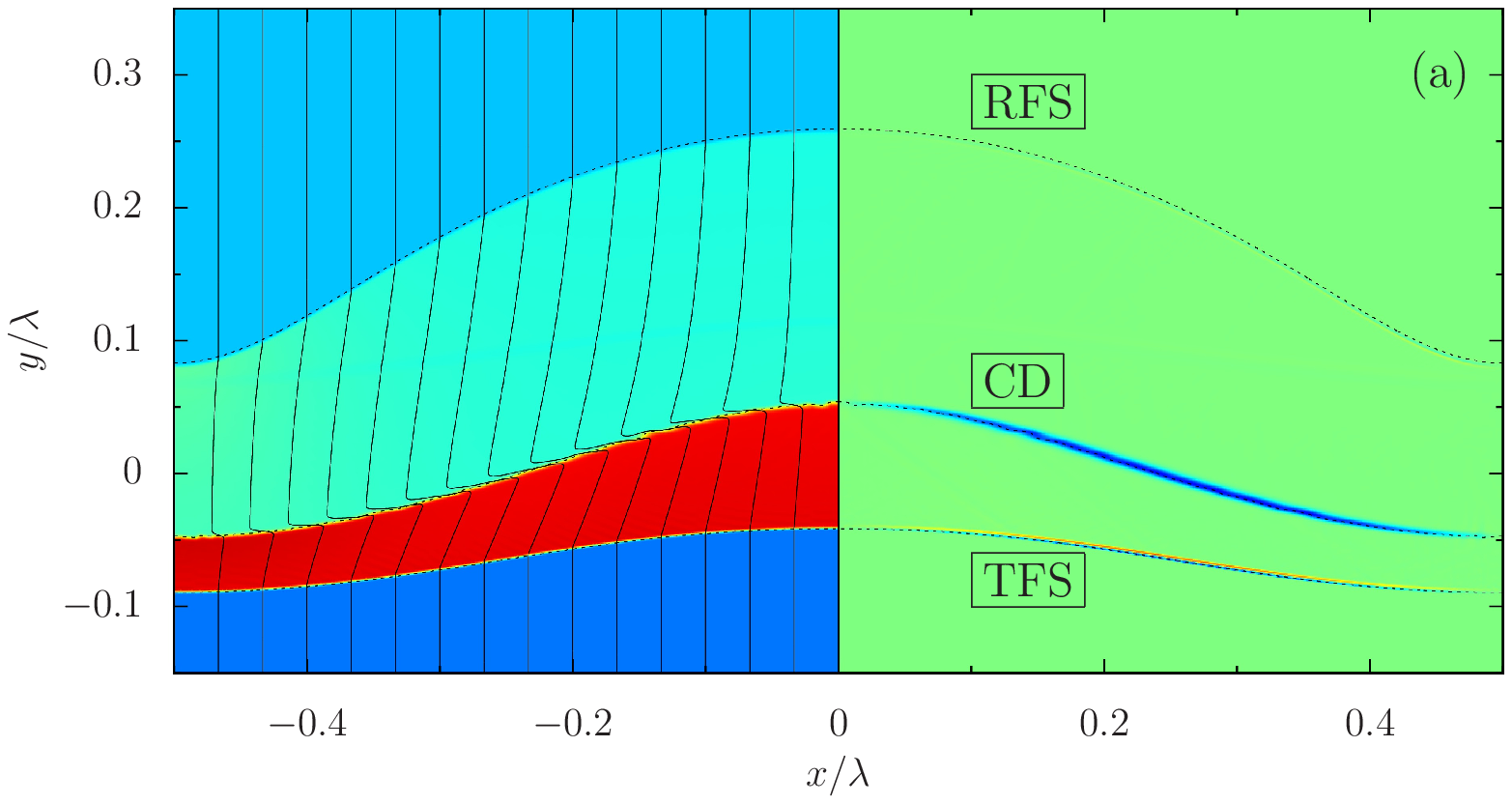} \\
\vspace{0.2cm}
\includegraphics[scale=0.54,clip]{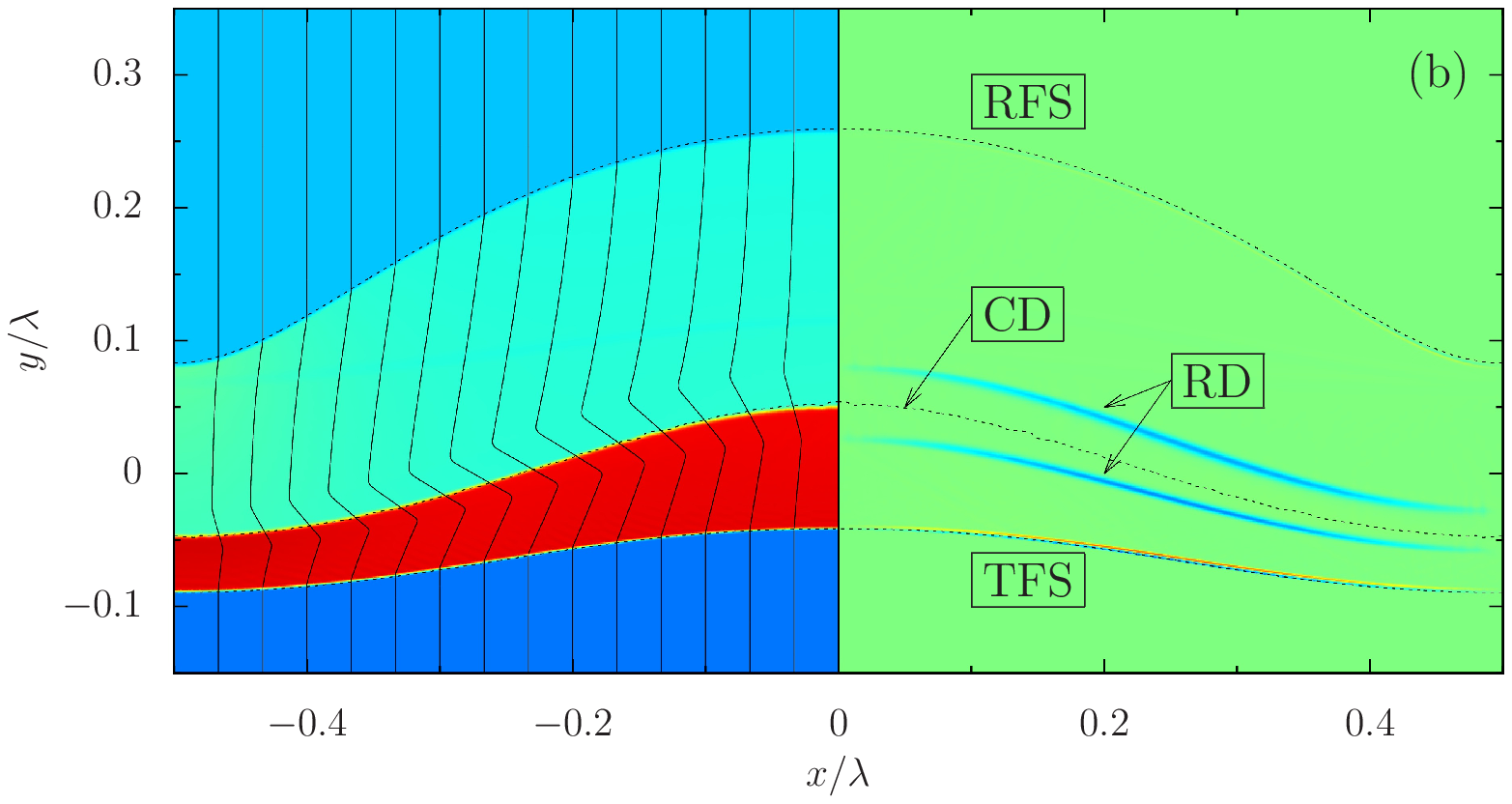}
\caption{
(a) 
Spatial distributions of the density (left panel) and the vorticity
(right panel) near the contact discontinuity. 
These snapshots are taken just after the interaction with the incident
shock at $k v_{\rm lin} t = 0.2$.  
The magnetic field lines are also depicted over the density profile.  
The dotted curves denote the surfaces of the reflected fast shock
(RFS), contact discontinuity (CD), and transmitted fast shock (TFS), 
from top to bottom.
The model parameters are $M = 20$, $\psi_0/\lambda = 0.1$,
$\rho_2/\rho_1 = 3$, and $\beta_0 = 100$. 
The higher resolution $\Delta_x = \Delta_y = \lambda/1024$ is used for
this figure. 
(b) 
The same figure for a model with a stronger initial field with 
$\beta_0 = 0.1$. 
The other parameters are identical to the model in (a). 
The rotational discontinuities (RD) appear at the both sides of the CD
for this case. 
\label{fig2}}
\end{figure}

Figures~\ref{fig2}(a) and \ref{fig2}(b) show the zoomed-in views of
spatial distributions for the density, field lines, and vorticity.
The snapshot data are taken just after the interaction at $k v_{\rm
  lin} t = 0.2$. 
These figures are to demonstrate the comparison between a weak field
case $\beta_0 = 100$ [Fig.~\ref{fig2}(a)] and a strong field case
$\beta_0 = 0.1$ [Fig.~\ref{fig2}(b)].  
All the model parameters other than $\beta_0$ are identical, which are
$M = 20$, $\psi_0/\lambda = 0.1$, and $\rho_2/\rho_1 = 3$.

Three discontinuous surfaces can be recognized easily from the density
contrast, which are interpreted as the reflected fast shock, contact
discontinuity, and transmitted fast shock. 
At this early stage, the density distributions are almost the same in
the both cases. 
However, obvious differences have appeared later at the nonlinear
regime of the RMI.  

In fact, significant growth of a spike due to the RMI can be seen only
in the weak field case.    
For this case, the shape of the vortex sheet always evolves together
with the contact surface throughout the calculation.   
Despite of a serious kink of the field lines near the contact
discontinuity, the weak field cannot influence the fluid motions and
vorticity distribution. 
Therefore, the RMI grows nonlinearly in a similar manner as the
hydrodynamical cases.   

For the strong field case, on the other hand, the vorticity is
no longer associated with the contact discontinuity and split into two
oppositely propagating sheets. 
For this case, another discontinuity in between the fast shock and
contact surface can be identified by a kink of the field lines as well
as the location of the vortex sheet.
Because the density is continuous and the direction of the tangential
field is opposite across this discontinuity, the structure coincides
with the rotational discontinuity.
The propagation velocity of the rotational discontinuity corresponds
to the Alfv{\'e}n speed, and the extraction of the vorticity leads to
the suppression of the RMI.  


Even for the weak field case, the vortex sheet should be propagating
with the Alfv{\'e}n speed.
However, it would be too slow to make a difference in the evolutions
of the RMI.
For the model shown in Fig.~\ref{fig2}(a), the ratio of the Alfv{\'e}n
speed to the growth velocity is much smaller than unity.
When the Alfv{\'e}n speed becomes comparable to $v_{\rm lin}$ given by
Eq.~(\ref{eq:vlin}), then the growth of the RMI seems to be severely
reduced. 
It is inferred from this fact that the competition between the
Alfv{\'e}n speed and $v_{\rm lin}$ could be a controlling factor of
the MHD RMI.  

Here we introduce a condition that the Alfv{\'e}n speed exceeds the
growth velocity of the RMI, 
\begin{equation}
v_{A2}^{\ast} \gtrsim \alpha v_{\rm lin} \;,
\label{eq:cond}
\end{equation}
where the Alfv{\'e}n speed is represented by $v_{A2}^{\ast} = B_0 / (4
\pi \rho_2^{\ast})^{1/2}$ estimated in the heavy fluid 2.
The growth velocity at the nonlinear regime is assumed to be $\alpha
v_{\rm lin}$ where $\alpha$ is typically of the order of 0.1 based on
the direct numerical simulations \cite{dimonte10,sano12}. 

Then, the critical field strength $B_{\rm crit} \equiv (4 \pi
\rho_2^{\ast})^{1/2} \alpha v_{\rm lin}$ can be expressed in terms of
$\beta$ as 
\begin{equation}
\beta_{\rm crit} \equiv \frac{8 \pi P_0}{B_{\rm crit}^2} 
= \frac2{\gamma} \alpha^{-2} 
\left( \frac{v_{\rm lin}}{c_{s2}^{\ast}} \right)^{-2}
\left( \frac{P^{\ast}}{P_0} \right)^{-1} \;,
\label{eq:bcrit}
\end{equation}
where $c_{s2}^{\ast} \equiv (\gamma P^{\ast} / \rho_2^{\ast})^{1/2}$
is the postshock sound speed in the spike, and $P^{\ast}$ is the
postshock pressure at the interface. 
The critical $\beta$ given by Eq.~(\ref{eq:bcrit}) can be
evaluated by solving a Riemann problem relevant to a set of the
initial parameters $M$ and $\rho_2/\rho_1$.
Notice that $\beta_{\rm crit}$ is defined by the preshock pressure
$P_0$. 

The growth velocity $v_{\rm lin}$ is roughly proportional to the
incident shock velocity $U_i$, or the Mach number $M$.
Then the critical strength $B_{\rm crit}$ will be proportional to $M$,
because the postshock density $\rho_2^{\ast}$ is almost constant in
the strong shock limit. 
The sound speed $c_{s2}^{\ast}$ in Eq.~(\ref{eq:bcrit}) is typically
comparable to or an order of magnitude smaller than $v_{\rm lin}$. 
The ratio $c_{s2}^{\ast}/v_{\rm lin}$ has little dependence on the model
parameters, and then the size of $\beta_{\rm crit}$ is determined by 
the amplification factor of the pressure $P^{\ast} / P_0$. 


\begin{figure}
\includegraphics[scale=0.49,clip]{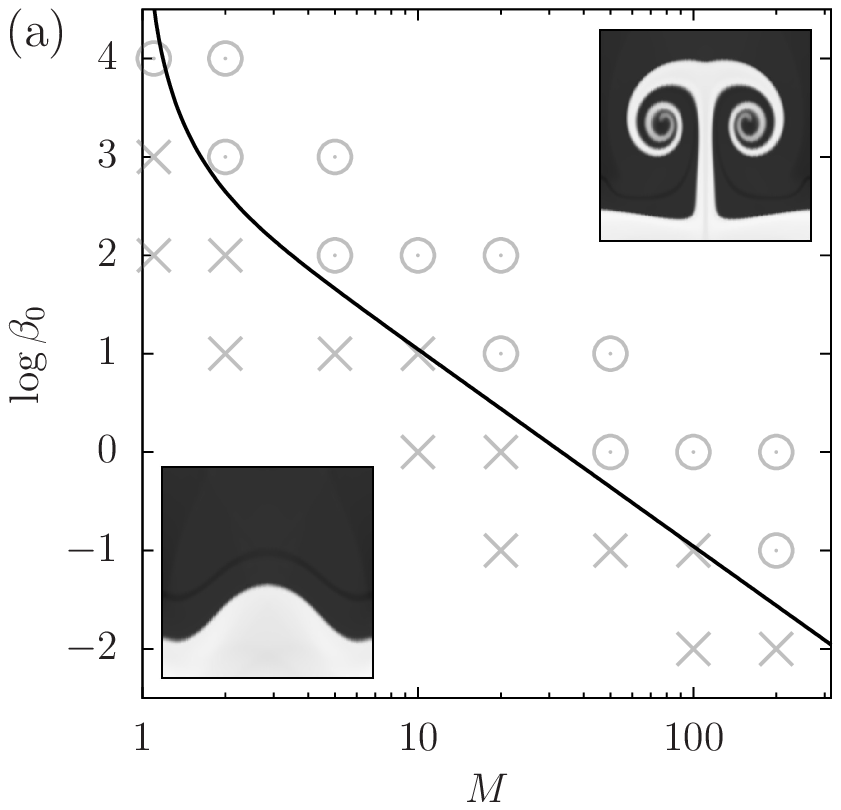}%
\hspace{0.2cm}\includegraphics[scale=0.49,clip]{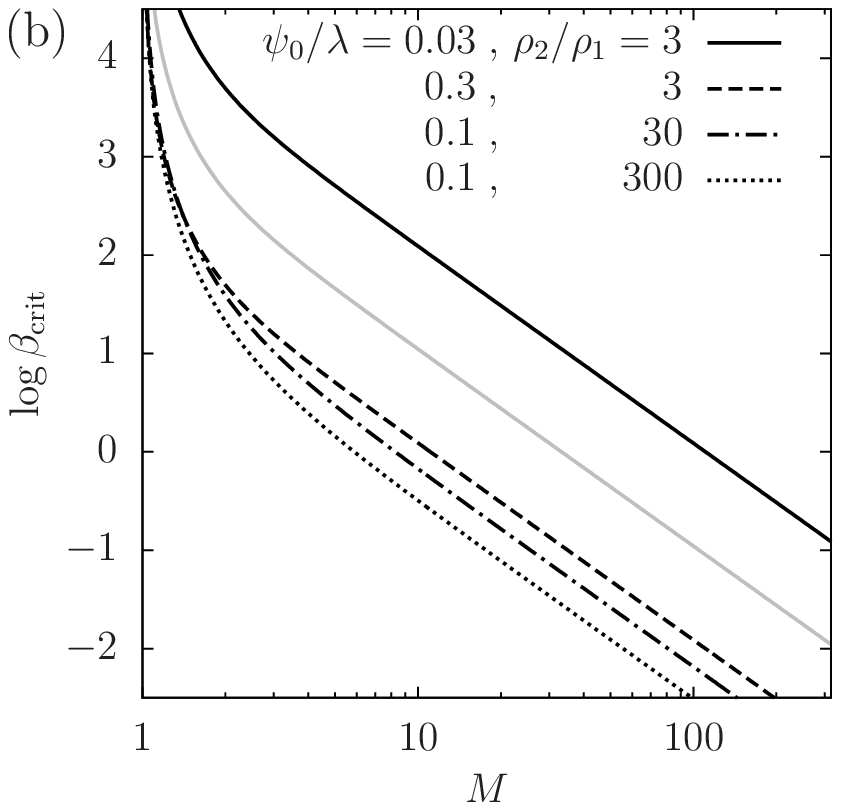}
\caption{
(a)
The critical field strength as a function of the Mach number of the
incident shock. 
The solid curve is for the fiducial case with $\psi_0/\lambda = 0.1$ and
$\rho_2/\rho_1 = 3$.  
The circles denote the unstable models that exhibit the nonlinear
growth of RMI in the numerical simulations, whereas the crosses stand
for the models which are stabilized by the ambient magnetic field.
Insets are typical results for the shape of the interface at $k
v_{\rm lin} t = 10$ in an unstable model with $\beta_0 = 100$
and $M = 20$ (top panel), and in a stabilized model with $\beta_0 =
0.1$ and $M = 20$ (bottom panel).
(b)
The critical field strength estimated form Eq.~(\ref{eq:bcrit}) for
various set of parameters $\psi_0/\lambda$ and $\rho_2/\rho_1$.
The parameter space above the critical curve corresponds to the
unstable regions for the MHD RMI.
For comparison, the critical curve for the fiducial case
($\psi_0/\lambda = 0.1$ and $\rho_2/\rho_1 = 3$) is also shown by the
gray curve. 
\label{fig3}}
\end{figure}

In Fig.~\ref{fig3}(a), the critical field strength is shown by the solid
curve as a function of the Mach number $M$.
As for the fiducial case, we choose $\psi_0/\lambda = 0.1$ and
$\rho_2/\rho_1 = 3$. 
As expected, the critical $\beta$ decreases dramatically as $M$
increases. 
This condition suggests that, for the strong shock cases ($M \gtrsim
30$), the RMI could occur under a very strong ambient field as
$\beta_0 \lesssim 1$. 
But if the shock is weak ($M \lesssim 3$), then the RMI will be suppressed
even by a weaker field of $\beta_0 \gtrsim 100$. 


In order to confirm the validity of this criterion, we performed
the direct numerical simulations of the RMI for various sets of
parameters for $M$, $\psi_0/\lambda$, $\rho_2/\rho_1$, and $\beta_0$.
Nonlinear outcomes of the RMI in each model are depicted by the gray
marks in Fig.~\ref{fig3}(a). 
The circles in this figure denote the models in which the nonlinear
growth of the RMI can be seen.
We define this condition that the growth velocity of the spike
continues to take positive values until at least $k v_{\rm lin} t = 10$. 
Actually, this is almost equivalent to the condition that the
magnetic field is amplified more than 10 times compared to the
initial value $B_0$ by this timescale ($k v_{\rm lin} t = 10$). 
The crosses, on the other hand, stand for the models where the RMI is
stabilized due to the existence of a magnetic field. 
In those models, neither the growth of mushroom-shaped
spike nor the field amplification can be realized.
As seen from Fig.~\ref{fig3}(a), the criterion given by
Eq.~(\ref{eq:bcrit}) is predicting the nonlinear results of
the numerical simulations with fairly good accuracy.  


The dependence of the critical $\beta$ on the other parameters is
shown by Fig.~\ref{fig3}(b).  
When the corrugation amplitude is larger (smaller), the growth
velocity becomes faster (slower).  
Then the critical curve shifts downward (upward) in the $M$-$\beta$
diagram.  
The larger density jump at the contact surface causes faster growth of
the RMI. 
For the cases of $\rho_2 / \rho_1 = 300$, the critical $\beta$ becomes
2 orders of magnitude smaller than that for $\rho_2 / \rho_1 = 3$.
All the critical curves in Fig.~\ref{fig3}(b) are reasonably
consistent with the simulation results, so that the criterion given by
Eq.~(\ref{eq:bcrit}) can be applicable for various situations of the RMI. 

The transverse field component $B_x$ also has the stabilizing effect
for the RMI through the Lorentz force \cite{cao08}.  
Our numerical simulations reveal that the critical strength 
$B_{\rm crit}$ for the perpendicular shock cases is quantitatively
similar to that for the parallel shock cases.
We have checked the convergence of the numerical results with
respect to the grid resolution.
The nonlinear behaviors of the RMI shown in Fig~\ref{fig3}(a) are
unchanged even with the quarter resolution $\Delta = \lambda/64$ and the
quadruple resolution $\Delta = \lambda/1024$.  

In summary, we have derived a formula of the critical field
strength for suppression of the RMI.
The obtained criterion is quite useful for the estimation of the MHD
effects on the RMI.
For the weak shock cases, for example, a weak field that is
dynamically unimportant as $\beta_0 \gg 1$ can reduce the growth of
the RMI significantly. 
When the incident shock is strong enough, on the
other hand, nonlinear growth of the RMI is allowed even when the
initial plasma $\beta$ is less than unity. 
Notice that the high Mach number $M \gg 10$ is characteristics of
the interstellar shocks driven by supernova explosions. 
For those cases, significant growth of the RMI can be
expected even when the preshocked gas is strongly magnetized as $\beta_0
\sim 1$. 
Then the magnetic field can be amplified locally near the interface up
to the order of the turbulent energy $B_{\max}^2 \sim \rho v_{\rm lin}^2$
\cite{sano12}. 
This amplification mechanism could explain the origin of the milligauss
field observed at young supernova remnants \cite{uchiyama07}.

The linear impulsive model for the MHD RMI \cite{wheatley05} shows
that the asymptotic amplitude of the interface tends to be
$\psi_{\infty} \approx \psi_0 ( 1 + V / v_{{\rm A}2}^{\ast} )$ when
$\rho_2^{\ast} > \rho_1^{\ast}$, where $V$ is the initial velocity of
the impulsively accelerated interface. 
Therefore, the suppression condition can be expressed as 
$V / v_{{\rm A}2}^{\ast} \ll 1$, which is qualitatively consistent
with our criterion.
The growth velocity at the nonlinear regime is assumed empirically as
$0.1 v_{\rm lin}$ in this analysis.  
Obviously the quantitative improvement of the criterion will be an
important next step.
Furthermore, the extension to three-dimensions is inevitable for the
studies of RMI \cite{long09}.
When the initial field direction is purely in the $z$-direction,
the RMI can always grow independent of the field strength 
within the two-dimensional approximation \cite{inoue12b}.
This feature should cause the asymmetric evolutions of MHD RMI, and
thus the three-dimensional analysis will be also an interesting subject
for our future work.  

\begin{acknowledgments}
Computations were carried out on SX-8R at the Cybermedia Center and
SX-9/B at the Institute of Laser Engineering of Osaka University. 
This research work is partly supported by results of HPCI Systems
Research Projects (Project ID hp120227).
\end{acknowledgments}

%





\end{document}